\documentclass[12pt]{article}
\usepackage[utf8]{inputenc}

\usepackage{amsmath}
\usepackage{amsfonts}
\usepackage{amssymb}
\usepackage{makeidx}
\usepackage{graphicx}
\usepackage{subcaption}
\usepackage{listings}

\usepackage{apacite}
\usepackage{natbib}

\usepackage{indentfirst}
\usepackage{caption}

\usepackage{ulem}
\usepackage{epsfig}

\usepackage{siunitx}
\usepackage{booktabs}
\newcolumntype{Y}{>{\centering\arraybackslash}X}

\usepackage{graphicx,psfrag,epsf}
\usepackage{enumitem}

\usepackage{url} 
\usepackage{newtxmath}  
\usepackage{setspace}
\usepackage{bbm}
\usepackage{dsfont}
\usepackage{newtxtext}      
\usepackage{url}

\usepackage{authblk}
\usepackage[pdfborder={0 0 0}]{hyperref}

\title{\bf Option Pricing in an Investment Risk-Return Setting}
\author[a]{Abootaleb Shirvani}
\author[b]{Frank J. Fabozzi}
\author[c]{Stoyan V. Stoyanov}

\begin{spacing}{.7}
\affil[a]{\small Department of Mathematics and Statistics, Texas Tech University\\
\url{abootaleb.shirvani@ttu.edu}}
\affil[b]{\small EDHEC Business School\\
\url{frank.fabozzi@edhec.edu}}
\affil[c]{\small College of Business, Stony Brook University\\
	\url{stoyan.stoyanov@stonybrook.edu}}
\end{spacing}

\begin{document}
\thispagestyle{plain}

\date{}

\maketitle
\begin{spacing}{1.00}
\noindent \textbf{Abstract}\ \ \ \ \ 	In this paper, we combine modern portfolio theory and option pricing theory so that a trader who takes a position in a European option contract and the underlying assets can construct an optimal portfolio such that at the moment of the contract's maturity the contract is perfectly hedged.  We derive both the optimal holdings in the underlying assets for the trader's optimal mean-variance portfolio and the amount of unhedged risk prior to maturity. Solutions assuming the cases where the price dynamics in the underlying assets follows discrete binomial price dynamics, continuous diffusions, stochastic volatility, volatility-of-volatility, and Merton-jump diffusion are derived.
\\
\\
\noindent \textbf{Keywords}\ \ \ \    Option pricing; mean-variance portfolio; binomial pricing trees; stochastic continuous diffusions; stochastic volatility; volatility-of-volatility; and Merton jump diffusions.
\end{spacing}

\begin{spacing}{1.00}
\newpage

\section{Introduction}

The pioneering work of \cite{Markowitz:1952} provides a framework for optimal portfolio construction of risky assets. Following the fundamental contributions of \cite{Black:1973} and \cite{Merton:1973} on option pricing, the mean-variance framework has been adopted to solve the problem related to option pricing such as optimal hedging of derivatives (see See \cite{Biagini:2000} and the review provided by \cite{Schweizer:2010}). Other applications of the mean-variance framework include optimal portfolios construction from derivative assets (see \cite{Jones:2006}, \cite{Eraker:2007}, \cite{Driessen:2013}, and \cite{Faias:2017}). In this paper we propose a method for constructing an optimal portfolio from a derivative contract and the underlying assets with the additional constraint that the derivative contract becomes perfectly hedged at the time of maturity. Our approach can be viewed as a combination of mean-variance analysis and option pricing theory.

We consider a trader who takes a short position in a European contingent claim (ECC) contract and takes a delta position in the underlying assets. The objective of the trader is to have a perfect hedge at the maturity of the ECC contract. Between the initiation of the ECC-contract and its maturity, the trader maximizes the portfolio of the underlying assets smoothly in time in a mean-variance framework choosing a time dependent risk-aversion coefficient. We calculate both the optimal holdings in the underlying assets for the trader's optimal mean-variance portfolio and the amount of unhedged risk prior to maturity. At maturity, the trader's portfolio provides a perfect hedge. 

Our empirical results in the binomial pricing model case indicate that when the call options are in-the-money, traders with the short position in a call option would like to guarantee a perfect hedge for the portfolio prior to the terminal date. The results show in the out-the-money call-option traders do not use the investment opportunity suggested in this paper. When the option is in-the-money, traders do use extensively the investment strategy offers.

We consider several cases in this paper. In the next section, we construct the trader's optimal portfolio assuming discrete binomial price dynamics for the underlying assets.  The case when the underlying assets follow continuous diffusions is provided in Section 3. The trader's optimal portfolio when the underlying assets follow a stochastic volatility model is analyzed in Section 4. In Section 5, we study the trader's optimal portfolio within the setting of a volatility-of-volatility model. The trader's optimal holdings in the underlying assets using the Merton jump-diffusion model is derived in Section 6. In Section 7 we provide concluding remarks.

\section{Binomial Pricing Trees with Optimal Quadratic Utility}

\noindent    We start with the following generalization of the Cox-Ingersoll-Ross (CRR) model as presented in \cite{Cox:1979}.  

\subsection{Option Pricing with Optimal Quadratic Utility  with following generalization of the Cox-Ingersoll-Ross model}
First, we consider one-step binomial pricing model (see \cite{Kim:2016}, KSRF hereafter). The asset price at $t=0\ \ $is denoted by $S_0$, representing the current known asset price. The option price on the asset at $t=0\ \ $is denoted by $f_0$. The option expiration time is $T>0.$ At $T$ , the asset price is denoted by  $S_T$, and the asset price can either move up to $S^u_T=uS_0$  for some $u>1,$ or down to $S^d_T=dS_0$, for some $d\in \left(0,1\right).$ The probability for the upward movement is $p\in \left(0,1\right)$, and  the probability for downward movement is $1-p.$ When the price of the asset price moves up to level  $S^u_T$, the final option payoff  is $f_T=f^u_T$: if the asset price moves down to $S^d_T$, the final option payoff is $f_T=f^d_T$.

Let's designate the trader as $\aleph .\ $ $\aleph $ is the option seller who takes a  $\mathrm{\Delta }$-position in the asset and i shorts the option contract. $\aleph $ forms a portfolio $P_0\mathrm{=}\mathrm{\Delta }S_0-f_0.$ At time $T,$ the value of the portfolio is$\ P_T$ and it can be either $P^u_T=\mathrm{\Delta }S^u_T-f^u_T$ with probability $p$, or  $P^d_T=\mathrm{\Delta }S^d_T-f^d_T$ with probability $1-p$. $\aleph$ choses a relative risk-averse constant $C\in (0,1]$, and would like to maximize the expected utility function  $\mathcal{U}\left(\ P_T\right)=\left(1-C\right)\mathbb{E}\left(P_T\right)-Cvar\left(P_T\right)$. One can choose different expected utility functions, such as  $\mathcal{U}\left(\ P_T\right)=\left(1-C\right)\mathbb{E}(P_T-rT)-Cvar(P_T)$, or $\mathcal{U}\left(\ P_T\right)=\frac{\mathbb{E}\left(P_T\right)-rT}{\sqrt{var(P_T)}}.$ The methods we employ in this paper can be readily extended to those two cases.

This leads to 
\begin{equation} \label{GrindEQ__1_} 
	\mathrm{\Delta }\mathrm{=}{\mathrm{\Delta }}^{\left(p\right)}_{\mathrm{C}}=\frac{\mathrm{pu+}\left(\mathrm{1-p}\right)\mathrm{d}}{\mathrm{2}\mathbb{R}\mathrm{p}\left(\mathrm{1-p}\right){\mathrm{S}}_0{\left(\mathrm{u-d}\right)}^{\mathrm{2}}}+{\mathrm{\Delta }}^{\mathrm{(r-n)}}, 
\end{equation} 
where $\mathbb{R}=\frac{\mathrm{C}}{\mathrm{1-C}}\in \left(0,\infty \right)$ denotes the absolute risk aversion constant, and                                                      
\begin{equation}
\label{GrindEQ__2_} 
{\mathrm{\Delta }}^{\mathrm{(r-n)}}=\frac{f^u_T-f^d_T}{S_0(u-d)}                                                               
\end{equation} 
\noindent is the $\mathrm{\Delta }$-position in the risk-neutral portfolio. If $C=1,$ $\mathbb{R}=\infty ,$ then $\aleph $'s portfolio exhibits minimal variance and the corresponding $\mathrm{\Delta }\mathrm{=}{\mathrm{\Delta }}^{\mathrm{min-var}}$ is equal to ${\mathrm{\Delta }}_{\mathrm{1}}={\mathrm{\Delta }}^{\mathrm{(r-n)}}$. If $C\downarrow 0,\mathbb{R}\mathrm{=}0,$ $\aleph $'s portfolio's risk explodes to infinity, and thus, $\mathrm{\Delta }\mathrm{=}{\mathrm{\Delta }}_0\uparrow \infty $, as well.  

Indeed, in one-period pricing, it is impossible to reconcile $\aleph$'s desire to maximize the portfolio utility function $\mathcal{U}\left(\ P_T\right)=\left(1-C\right)\mathbb{E}\left(P_T\right)-Cvar\left(P_T\right)$, and at the same time to be able to replicate the option-payoff at maturity. This is because the risk aversion parameter $C>0$ is fixed. Next, we will extend $\aleph$'s absolute risk aversion parameter ${\mathbb{R}}_{\tau }=\frac{{\mathrm{C}}_{\mathrm{\tauup }}}{\mathrm{1-}{\mathrm{C}}_{\mathrm{\tauup}}}\in \left(0,\infty \right),\tau =T-t $ We will use both notations  ${\mathbb{R}}_{\tau }=\frac{{\mathrm{C}}_{\mathrm{\tauup }}}{\mathrm{1}\mathrm{-}{\mathrm{C}}_{\mathrm{\tauup }}}\in \left(0,\infty \right),\tau =T-t\in \left[0,T\right]$ and   ${\mathbb{R}}_t=\frac{{\mathrm{C}}_{\mathrm{t}}}{\mathrm{1-}{\mathrm{C}}_t}\in \left(0,\infty \right),t=T-\tau \in [0,T]$ to increase to infinity as the time to option's maturity  $\tau \downarrow 0$. That is, ${\mathrm{lim}}_{\tau \downarrow 0}{\mathrm{C}}_{\mathrm{\tauup }}=1$, and  In this way, $\aleph$ can realize a portfolio gain in $\left[0,T\right)$ while guaranteeing a fully replication of ${\aleph}$'s liability in the short position at the option's maturity $T$.   

To this end, we extend the one-step pricing model to the binomial model in the setting proposed by KSRF. We assume that the asset is traded in instances ${\mathrm{t}}_{\mathrm{k}}=kh,\ k=0,1.,,,n,nh=T,\ h\downarrow 0.$ The price dynamics is given by
\begin{equation} \label{GrindEQ__3_} 
	{\mathrm{S}}_{{\mathrm{t}}_{\mathrm{k+1}}}={\mathrm{S}}_{{\mathrm{t}}_{\mathrm{k}}}\left\{ \begin{array}{ll}
		1+\mu h+\sqrt{\frac{1-p_h}{p_h}}\sigma \sqrt{h}\ ,\mathrm{\ with\ probability}\ p_h\, ,  \\ 
		1+\mu h-\sqrt{\frac{p_h}{1-p_h}}\sigma \sqrt{h}\ ,\mathrm{\ with\ probability}\ 1-p_h \end{array}
	\right. 
\end{equation} 
for $k=0,1.,,,n-1$ , where $p_h\in \left(0,1\right)$ is a given probability for the upward movement of the asset price in any time interval $\left({\mathrm{t}}_{\mathrm{k}},{\mathrm{t}}_{\mathrm{k+1}}\right]$. In \eqref{GrindEQ__3_}, $\mu >r>0$ is the instantaneous asset price mean return, where $r$ is the instantaneous riskless rate, and $\sigma >0$ is the asset-volatility. $\aleph $'s portfolio at ${\mathrm{t}}_{\mathrm{k}}$ is then  $P_{{\mathrm{C}}_{{\mathrm{\tauup }}_{\mathrm{k}}},{\mathrm{t}}_{\mathrm{k}}}={\mathrm{\Delta }}^{\left(p_h\right)}_{{\mathrm{C}}_{{\mathrm{\tauup }}_{\mathrm{k}}},{\mathrm{t}}_{\mathrm{k}}}S_{{\mathrm{t}}_{\mathrm{k}}}-f_{{\mathrm{t}}_{\mathrm{k}}},\ $where $f_{t_k}$ is the ECC's option's value at $t_k$. Here, $f_{t_k}$ is the fair (risk-neutral) option value as derived in the KSRF framework.\footnote{ If $p_h$ is chosen to be $p_h=\frac{1}{2}+\frac{\mu -\frac{{\sigma }^2}{2}}{2\sigma }\sqrt{h},h\downarrow 0$, the KSRFmodel is identical to the CRR model for option valuation using the binomial pricing tree. In KSRF, $p_h$ can have any value in $\left(0,1\right)$. Typically, $p_h$ is estimated from historical data on the number of times the asset's price moves up and the number of times it moves down.} Then, according to \eqref{GrindEQ__1_} and \eqref{GrindEQ__2_}, the $\mathrm{\Delta }$-position in the asset at ${\mathrm{t}}_{\mathrm{k}}\ $is given by
\begin{equation} 
\label{GrindEQ__4_}
{\mathrm{\Delta }}^{\left(p_h\right)}_{{\mathrm{C}}_{{\mathrm{\tauup }}_{\mathrm{k}}},{\mathrm{t}}_{\mathrm{k}}}=\frac{\mu }{2{{\mathbb{R}}_{{\mathrm{\tauup }}_{\mathrm{k}}}\mathrm{S}}_{{\mathrm{t}}_{\mathrm{k}}}{\sigma }^2}+\frac{f^u_{{\mathrm{t}}_{\mathrm{k+1}}}-f^d_{{\mathrm{t}}_{\mathrm{k+1}}}}{S_{{\mathrm{t}}_{\mathrm{k}}}\sigma \sqrt{h}}\sqrt{p_h\left(1-p_h\right)}
\end{equation}
where ${\mathbb{R}}_{{\mathrm{\tauup }}_{\mathrm{k}}}=\frac{{\mathrm{C}}_{{\mathrm{\tauup }}_{\mathrm{k}}}}{1-{\mathrm{C}}_{{\mathrm{\tauup }}_{\mathrm{k}}}}\in \left(0,\infty \right),{\tau }_k=T-t_k=\left(n-k\right)h$. The difference between the investment $\mathrm{\Delta }$-position ${\mathrm{\Delta }}^{\left(p_h\right)}_{\mathrm{C,}{\mathrm{t}}_{\mathrm{k}}}\ $and the risk-neutral $\mathrm{\Delta }$-position ${\mathrm{\Delta }}^{\left(p_h\right)}_{0,{\mathrm{t}}_{\mathrm{k}}}$ is determined by

\begin{equation}
\label{GrindEQ__5_}
{\mathrm{\Delta }}^{\left(p_h\right)}_{{\mathrm{C}}_{{\mathrm{\tauup }}_{\mathrm{k}}},{\mathrm{t}}_{\mathrm{k}}}\mathrm{\ }=\frac{\mu }{2{\mathbb{R}}_{{\mathrm{\tauup }}_{\mathrm{k}}}{\mathrm{S}}_{{\mathrm{t}}_{\mathrm{k}}}{\sigma }^2}+{\mathrm{\Delta }}^{\left(r-n\right)}_{\mathrm{1,}{\mathrm{t}}_{\mathrm{k}}}.
\end{equation}

Now, $\aleph $ (in view of the risk-return preferences) can choose a non-decreasing continuous function  ${\psi }^{\left(\aleph \right)}\left(\tau \right)\ge 0,\ \tau \in [0,T]$ with ${\psi }^{\left(\aleph \right)}\left(0\right)=0$. If $\aleph $ would like to guarantee a perfect hedge for the portfolio when time-to-maturity $\ {\tau }_k=T-t_k$ is close to zero (i.e., $\ k\ $ is close to $n)$, $\aleph $ can choose 
\begin{equation}
\label{GrindEQ__6_}
{\psi }^{\left(\aleph ,a,\gamma \right)}\left(\tau \right)=
\left\{ \begin{array}{ll}
0\, , \ for\ \ \ \tau \ \in [0,a]\ \  \\ 
a-ae^{-\gamma \left(\tau -a\right)},\ for\ \ \tau \in [a,\infty ) 
\end{array}\right.
\end{equation}

with parameters $a>0$ and  $\gamma >0$.  We choose one parametric family
\begin{equation}
\label{GrindEQ__7_}
{\psi }^{\left(\aleph ,\gamma \right)}\left(\tau \right)=\gamma -\gamma e^{-\gamma \frac{\tau }{T}}.
\end{equation}

 If $\aleph $ would like a very small unhedged risk prior to the option's maturity, he should choose $\boldsymbol{\gamma }$ close to zero. We call parameter $\gamma >0$ as the risk aversion intensity in the option's short position that $\mathrm{\aleph }$ is taking in the option contract\textbf{.} Having chosen $\gamma >0,$ $\aleph $ computes the absolute risk-aversion constant ${\mathbb{R}}_{{\mathrm{\tauup }}_{\mathrm{k}}}=\ {\mathbb{R}}_{{\mathrm{\tauup }}_{\mathrm{k}}}\left({\mathrm{S}}_{{\mathrm{t}}_{\mathrm{k}}},\mu ,\sigma ,\gamma \right)=\frac{C_{{\mathrm{\tauup }}_{\mathrm{k}}}\left({\mathrm{S}}_{{\mathrm{t}}_{\mathrm{k}}},\mu ,\sigma ,\gamma \right)}{1-C_{{\mathrm{\tauup }}_{\mathrm{k}}}\left({\mathrm{S}}_{{\mathrm{t}}_{\mathrm{k}}},\mu ,\sigma ,\gamma \right)},$ in every time step $\left[t_k,t_{k+1}\right),\ k=0,\dots ,n-1,\ $ as the solution of  $\frac{\mu }{2{\mathbb{R}}_{{\mathrm{\tauup }}_{\mathrm{k}}}{\mathrm{S}}_{{\mathrm{t}}_{\mathrm{k}}}{\sigma }^2}={\psi }^{\left(\aleph ,\gamma \right)}\left({\tau }_k\right).$ Thus,

\begin{equation}
\label{GrindEQ__8_}
{\mathbb{R}}_{{\mathrm{\tauup }}_{\mathrm{k}}}=\ {\mathbb{R}}_{{\mathrm{\tauup }}_{\mathrm{k}}}\left({\mathrm{S}}_{{\mathrm{t}}_{\mathrm{k}}},\mu ,\sigma ,\gamma \right)=\frac{\mu }{2{\psi }^{\left(\aleph ,,\gamma \right)}\left({\tau }_k\right){\mathrm{S}}_{{\mathrm{t}}_{\mathrm{k}}}{\sigma }^2}.
\end{equation}

Having chosen the absolute risk-aversion constant according to \eqref{GrindEQ__7_}, $\aleph \ $choses the delta-position in the asset, namely ${\mathrm{\Delta }}^{\left(p_h\right)}_{{\mathrm{C}}_{{\mathrm{\tauup }}_{\mathrm{k}}},{\mathrm{t}}_{\mathrm{k}}}$, according to \eqref{GrindEQ__5_}. That is,
\begin{equation}
 \label{GrindEQ__9_} 	
 {\mathrm{\Delta }}^{\left(p_h\right)}_{{\mathrm{C}}_{{\mathrm{\tauup }}_{\mathrm{k}}},{\mathrm{t}}_{\mathrm{k}}}\mathrm{\ }={\psi }^{\left(\aleph ,\gamma \right)}\left({\tau }_k\right)+{\mathrm{\Delta }}^{\left(r-n\right)}_{\mathrm{1,}{\mathrm{t}}_{\mathrm{k}}} 
\end{equation} 
and ${\psi }^{\left(\aleph ,,\gamma \right)}\left({\tau }_k\right)>0$ vanishes at ${\tau }_n=0.$ At the last time interval $[T-h,T)\ {\tau }_{n-1}=n$, ${\aleph}$'s delta position ${\mathrm{\Delta }}^{\left(p_h\right)}_{{\mathrm{C}}_{{\mathrm{\tauup }}_{\mathrm{n-1}}},{\mathrm{t}}_{\mathrm{k}}}\mathrm{\ }={\psi }^{\left(\aleph ,\gamma \right)}\left(h\right)+{\mathrm{\Delta }}^{\left(r-n\right)}_{\mathrm{1,}{\mathrm{t}}_{\mathrm{k}}},$ where ${\psi }^{\left(\aleph ,\gamma \right)}\left(h\right)\sim {\gamma }^2\frac{h}{T}$ as $h\downarrow 0$. That is, as $h\downarrow 0$, ${\aleph}$'s portfolio becomes asymptotically riskless, guaranteeing (asymptotically, as the time-step $h\downarrow 0$) that $\aleph $ hedges the short position asymptotically at the terminal time $T$. If $\aleph$ would like to have a perfect hedge prior to the terminal time, he can choose \eqref{GrindEQ__6_} instead of \eqref{GrindEQ__7_}.  Prior to the terminal time $T$, ${\aleph }$'s portfolio 
\begin{equation}
\label{GrindEQ__10_}
P_{C_{{\tau }_k},t_k}={\mathrm{\Delta }}^{\left(p_h\right)}_{C_{{\tau }_k},t_k}S_{t_k}-f_{t_k}=\ {\psi }^{\left(\aleph ,\gamma \right)}\left({\tau }_k\right)S_{t_k}+{\mathrm{\Delta }}^{\left(r-n\right)}_{1,t_k}S_{t_k}-f_{t_k}{\mathrm{\Delta }}^{\left(r-n\right)}_{1,t_k}.
\end{equation}

Thus, as $h\downarrow 0,$  

\begin{equation}
\label{Profit}
P_{{\mathrm{C}}_{{\mathrm{\tauup }}_{\mathrm{k+1}}},{\mathrm{t}}_{\mathrm{k+1}}}-P_{{\mathrm{C}}_{{\mathrm{\tauup }}_{\mathrm{k}}},{\mathrm{t}}_{\mathrm{k}}}=rhP_{{\mathrm{C}}_{{\mathrm{\tauup }}_{\mathrm{k}}},{\mathrm{t}}_{\mathrm{k}}}+U^{\left(\aleph ,\gamma \right)}\left({\tau }_k\right),k=1,\dots ,n-1,
\end{equation}
where the value of the unhedged risk is given by 
\begin{equation}
\label{residual}
U^{\left(\aleph ,\gamma \right)}\left({\tau }_k\right)={\psi }^{\left(\aleph ,\gamma \right)}\left({\tau }_k\right)(\mu {-r)S}_{t_k}h+{\psi }^{\left(\aleph ,\gamma \right)}\left({\tau }_k\right)\sigma S_{t_k}\sqrt{h}{\mathcal{N}}_k
\end{equation}
with ${\mathcal{N}}_k,k=1,\dots ,n$ being a sequence of independent standard normal random variables, and 
\noindent ${\psi }^{\left(\aleph ,\gamma \right)}\left({\tau }_k\right)={\psi }^{\left(\aleph ,\gamma \right)}\left(kh\right)=\gamma -\gamma e^{-\gamma \frac{k}{n}h}\sim {\gamma }^2\frac{k}{n}h,\ k=0,1,...,n\ \ $as $h\downarrow 0$.

\subsection{Empirical Analysis}

Here we apply the new hedging method we proposed to explain
and evaluate the trader's investment opportunity for having a perfect hedge when traders take short positions in an option. We calculate the trader's optimal holding in the underlying asset for options with different times to maturity and strike prices. 

To this end, we first estimated $p_h$, the number of times that the asset's price moves up, from historical data. Then, using the price dynamics proposed by KSRF in \eqref{GrindEQ__3_}, we obtained $S^d_t$, the option payoff, $f_t$, and the $\mathrm{\Delta }$-position in the risk-neutral portfolio at $t_k$, $k=0,1,...,n$. For different values of $\gamma \in [0,10]$, denoted by $\gamma_{initial}$, we compute $P_{C_{{\tau }_k},t_k}$ and $P_{C_{{\tau }_{k+1}},t_{k+1}}$ from \eqref{GrindEQ__10_}, and consequently obtain $U^{\left(\aleph ,\gamma \right)}\left({\tau }_k\right)$ from \eqref{GrindEQ__11_} for $k=1,\dots ,n-1$. Then, we fit the normal distribution to $U^{\left(\aleph ,\gamma \right)}\left({\tau }_k\right)$ to estimate  ${\psi }^{\left(\aleph ,\gamma \right)}$ and $\gamma$, denoted by $\gamma_{end}$. If the initial $\gamma_{initial}$ is not equal to end $\gamma_{end}$, the procedure is repeated until a satisfactory solution is obtained ($\gamma_{initial}=\gamma_{end}$). 

We apply the methods described above to compute the optimal value of  $\gamma$ and ${\psi }^{\left(\aleph ,\gamma \right)}$. In our analysis, we selected a broad-based market index, the S\&P 500\footnote{See \url{https://us.spdrs.com/en/etf/spdr-sp-500-etf-SPY}.}, as measured by the SPDR S\&P 500  which is an exchange-traded fund  as the proxy for the aggregate stock market. We use the 10-year Treasury yield\footnote{\url{https://www.treasury.gov/resource-center/data-chart-center/interest-rates/Pages/TextView.aspx?data=yieldYear\&year=2019.}} as a proxy for $r$. The database covers the period from January 2015 to November 2019. There were 1,217 observations collected from \textit{Yahoo Finance}.

\begin{figure}[t!]
	\centering
	\includegraphics[scale=0.3]{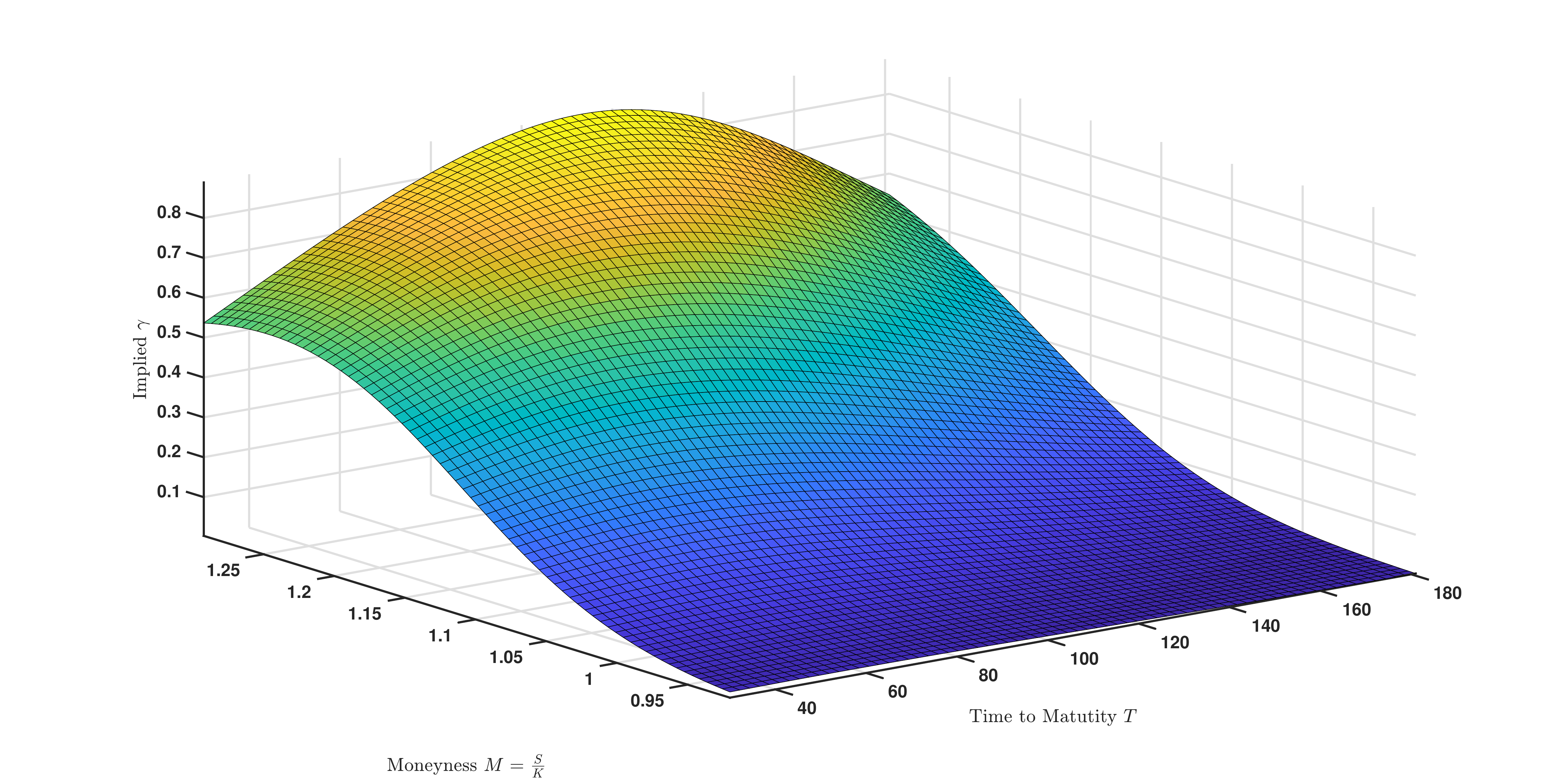}
	\caption{Plot of the risk aversion intensity $\gamma$  against time to maturity and moneyness.}
	\label{figur:gamma}
\end{figure}

The surface for the optimal value of $\gamma$ is graphed against both a standard measure of “moneyness” and time to maturity (in days) in Figure  \ref{figur:gamma}.
The Figure indicates that at each maturity, the optimal value for $\gamma$ increase as moneyness increases.  Where the moneyness varies in (0, 0.95), the surface is flat at point $0$, indicating traders with a short position in the option disregard the investment opportunity as suggested in this section.  Where the moneyness is greater than $1$, in-the-money options, traders do use the investment opportunity extensively.

\begin{figure}[htb!]
	\centering
	\includegraphics[scale=0.3]{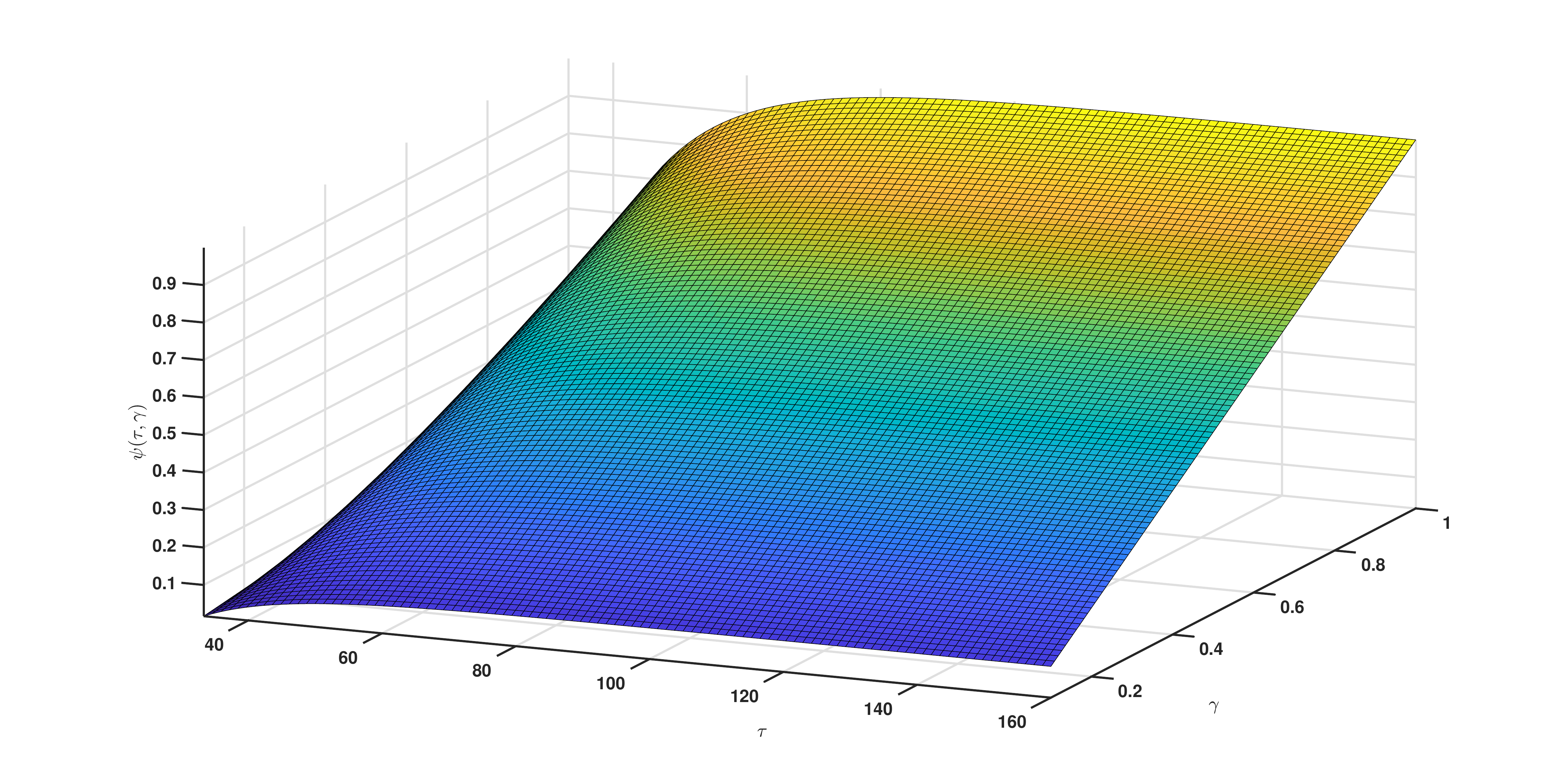}
	\caption{Plot of ${\psi }^{\left(\aleph ,\gamma \right)}\left(\tau \right)$, with terminal time $T=160 \, \left( \text{day}\right) , \tau \in \left[30,160\right],\gamma \in (0,1]$.}
	\label{Sai}
\end{figure} 

Figure  \ref{Sai} shows the surface of  ${\psi }^{\left(\aleph ,\gamma \right)}\left(\tau \right)=\gamma -\gamma e^{-\gamma \frac{\tau }{T}}$, against both  the risk-aversion intensity, $\gamma$, and time to maturity $\tau$. It illustrates the steepness of ${\psi }^{\left(\aleph ,\gamma \right)}\left(\tau \right)$ as $\tau$ converges to $0$ (i.e., $t\uparrow T$)  and the flatness of ${\psi }^{\left(\aleph ,\gamma \right)}\left(\tau \right)$ as time to maturity $\tau$ converges to $T$ (i.e., $t\downarrow 0$). If $\gamma$ is close to zero (i.e., when the option is out-the-money), traders with a short option position want to have a small unhedged risk prior to the option’s maturity. Conversely, when $\gamma$ increases from $0.1$ to $1$, traders would like to guarantee a perfect hedge for the portfolio prior to the terminal time by increasing the value of  $\psi$. Finally, we note that when $\tau$ is close to $T$, the steepness of ${\psi }^{\left(\aleph ,\gamma \right)}\left(\tau \right)$  increases when $\gamma$ increases. It again indicates that traders use the investment opportunity when the time to maturity is close to $0$.

\section{Option Pricing with Optimal Quadratic Utility when the Underlying Asset Price follows Continuous Diffusion}

\noindent Suppose the asset price follows continuous diffusion; that is,
\begin{equation} 
\label{GrindEQ__11_} 
	dS_t={\mu }_tS_tdt+{\sigma }_tS_tdB\left(t\right),\ S_0>0,\ t\in \left[0,T\right],   {\mu }_t>r_t>0,  {\sigma }_t>0, 
\end{equation} 
where $B\left(t\right),\ t\in \left[0,T\right],$ is a Brownian motion on the natural stochastic basis $\left(\mathrm{\Omega },\mathbb{F}=\left({\mathcal{F}}_t,\ t\in \left[0,T\right]\right)\mathrm{,}\mathrm{\ }\mathbb{P}\right),$ with filtration $\mathbb{F}$ generated by the Brownian motion $B\left(t\right),\ t\in \left[0,T\right].$ In \eqref{GrindEQ__11_}, ${\mu }_t$ is the instantaneous mean asset return at time  $t$, $\ {\sigma }_t$ is the asset volatility and $r_t$ is the riskless rate\footnote{ The processes  ${\mu }_t,\ {\sigma }_t$ and $r_t$ defined on $\left(\mathrm{\Omega },\mathbb{F}=\left({\mathcal{F}}_t,\ t\in \left[0,T\right]\right)\mathrm{,}\mathrm{\ }\mathbb{P}\right)$, satisfy the usual regularity conditions, see \citet[chapter 6]{Duffie:2001}.}. Proceeding in the same fashion as in Section 2, we assume that, at time $t\in \left[0,T\right),\ $ $\aleph $ forms a self-financing portfolio $P^{\left({\mathrm{C}}_{\mathrm{t}}\right)}_t\mathrm{=}{\mathrm{\Delta }}_{{\mathrm{C}}_{\mathrm{t}}}\left(t\right)S_t-f_t,\ $ where $f_t=V\left(S_t,t\right)$is the fair price of the option at time $t$, and thus $V\left(x,t\right),\ x>0,t\in \left[0,T\right),$ satisfying the Black-Scholes-Merton partial differential equation (BSM PDE) given by:
\[\frac{\partial V\left(x,t\right)}{\partial t}+r_tx\frac{\partial V\left(x,t\right)}{\partial x}-r_tV\left(x,t\right)+\frac{1}{2}{\sigma }^2_tx^2V{\left(x,t\right)}^2=0,\] 
with boundary condition $V\left(x,T\right)=G(x)$, where $f_T=V\left(S_T,T\right)=G(S_T)$ is the option's terminal value. Then the portfolio dynamics is given by

\begin{equation} 
\label{GrindEQ__12_} 
\begin{array}{ccc}
dP^{\left({\mathrm{C}}_{\mathrm{t}}\right)}_t={\mathrm{\Delta }}_{{\mathrm{C}}_{\mathrm{t}}}\left(t\right)dS_t-df_t =
\\\left({\mathrm{\Delta }}_{{\mathrm{C}}_{\mathrm{t}}}\left(t\right){\mu }_tS_t-\frac{\partial V\left(x,t\right)}{\partial t}-{\mu }_tS_t\frac{\partial V\left(x,t\right)}{\partial x}-\frac{1}{2}{\sigma }^2_tS^2_t\frac{{\partial }^2V\left(x,t\right)}{\partial x^2}\right)dt
\\
+\left({\mathrm{\Delta }}_{{\mathrm{C}}_{\mathrm{t}}}\left(t\right)-\frac{\partial V\left(x,t\right)}{\partial x}\right){\sigma }_tS_tdB\left(t\right) 
\end{array}
\end{equation} 

$\aleph \ $choses a dynamic relative risk averse parameter $C_t\in \left(0,1\right]$, and would like to maximize the conditional instantaneous expected utility function ${\mathcal{U}}_t\left(dP^{\left(C_t\right)}_t\right)$ determined by  ${\mathcal{U}}_t\left(dP^{\left(C_t\right)}_t\right)dt=\left(1-C_t\right){\mathbb{E}}_t(dP^{\left(C_t\right)}_t)-C_tvar_t({dP}^{\left(C_t\right)}_t)$, where ${\mathbb{E}}_t\left({dP}^{\left(C_t\right)}_t\right)=\ \mathbb{E}({dP^{\left(C_t\right)}_t}/{{\mathcal{F}}_t})$ and ${var}_t\left({dP}^{\left(C_t\right)}_t\right)=\ var({dP^{\left(C_t\right)}_t}/{{\mathcal{F}}_t})$ are the condition mean and conditional variance of ${dP}_t.$ From \eqref{GrindEQ__12_}, it follows that
\[{\mathbb{E}}_t\left({dP}^{\left(C_t\right)}_t\right)=\left({{\mathrm{\Delta }}_{\mathrm{C}}}_t\left(t\right){\mu }_tS_t-\frac{\partial V\left(x,t\right)}{\partial t}-{\mu }_tS_t\frac{\partial V\left(x,t\right)}{\partial x}-\frac{1}{2}{\sigma }^2_tS^2_t\frac{{\partial }^2V\left(x,t\right)}{\partial x^2}\right)dt,\] 
and ${var}_t\left({dP}^{\left(C_t\right)}_t\right)={{\sigma }^2_tS^2_t\left({\mathrm{\Delta }}_{{\mathrm{C}}_{\mathrm{t}}}\left(t\right)-\frac{\partial V\left(x,t\right)}{\partial x}\right)}^2dt$. Thus, 
\[{\mathcal{U}}_t\left(dP^{\left(C_t\right)}_t\right)=\left(1-C_t\right)\left({\mathrm{\Delta }}_{{\mathrm{C}}_{\mathrm{t}}}\left(t\right){\mu }_tS_t-\frac{\partial V\left(x,t\right)}{\partial t}-{\mu }_tS_t\frac{\partial V\left(x,t\right)}{\partial x}-\frac{1}{2}{\sigma }^2_tS^2_t\frac{{\partial }^2V\left(x,t\right)}{\partial x^2}\right)-\] 
\[-C_t{\sigma }^2_tS^2_t{\left({{\mathrm{\Delta }}_{\mathrm{C}}}_t\left(t\right)-\frac{\partial V\left(x,t\right)}{\partial x}\right)}^2.\] 
Then, the optimal ${{\mathrm{\Delta }}_{\mathrm{C}}}_t\left(t\right)$ is given by
\begin{equation} \label{GrindEQ__13_} 
 {\mathrm{\Delta }}_{{\mathrm{C}}_{\mathrm{t}}}\left(t\right)=\frac{{\mu }_t}{2{\mathbb{R}}_t{{\sigma }^2_tS}_t}+\mathrm{\ }{\mathrm{\Delta }}^{\left(\mathrm{r-n}\right)}\left(t\right), 
\end{equation} 
where ${\mathbb{R}}_t=\frac{{\mathrm{C}}_{\mathrm{t}}}{\mathrm{1-}{\mathrm{C}}_{\mathrm{t}}},\ $and  ${\mathrm{\Delta }}^{\left(\mathrm{r-n}\right)}\left(t\right)=\frac{\partial V\left(S_t,t\right)}{\partial x}={\mathrm{\Delta }}_{\mathrm{1}}\left(t\right)$ is the risk-neutral $\mathrm{\Delta }$-position.\footnote{As expected, when ${\mu }_t=\mu ,\ {\sigma }_t=\sigma ,r_t=r$, (13) represents the continuous version of (5).}  Now, as in \eqref{GrindEQ__7_} and \eqref{GrindEQ__8_} we define

\begin{equation} \label{GrindEQ__14_}  
{\mathbb{R}}_{\tau }={\mathbb{R}}_{\tau }\left({\mathrm{S}}_{\mathrm{t}},\mu ,\sigma ,\gamma \right)=\frac{{\mu }_t}{2{\psi }^{\left(\aleph ,\gamma \right)}\left(\tau \right){\mathrm{S}}_t{\sigma }^2_t},\ \ \ C_t=C_t\left({\mathrm{S}}_{{\mathrm{t}}_{\mathrm{k}}},\mu ,\sigma ,\gamma \right)=\frac{{\mathbb{R}}_{\tau }}{1+{\mathbb{R}}_{\tau }},
\end{equation}
where $\ {\psi }^{\left(\aleph ,\gamma \right)}\left(\tau \right)=\gamma -\gamma e^{-\gamma \frac{\tau }{T}}$ $=\frac{{\mu }_t}{2{\mathbb{R}}_{\tau }{\mathrm{S}}_t{\sigma }^2_t}$ and $\tau =T-t$. Thus, from \eqref{GrindEQ__13_},  it follows that   $\ {\mathrm{\Delta }}_{{\mathrm{C}}_{\mathrm{t}}}\left(t\right)={\psi }^{\left(\aleph ,\gamma \right)}\left(\tau \right)\mathrm{\ }{\mathrm{\Delta }}^{\left(\mathrm{r-n}\right)}\left(t\right)\sim \mathrm{\ }{\mathrm{\Delta }}^{\left(\mathrm{r-n}\right)}\left(t\right)$  $\mathrm{\ }{\mathrm{\Delta }}^{\left(\mathrm{r-n}\right)}\left(t\right)\sim \mathrm{\ }{\mathrm{\Delta }}^{\left(\mathrm{r-n}\right)}\left(t\right)$as $t$ approaches the terminal time $T,\ $asymptotically, as $t\uparrow T,$ $\aleph $'s portfolio$\mathrm{\ }P^{\left({\mathrm{C}}_{\mathrm{t}}\right)}_t\mathrm{=}{\mathrm{\Delta }}_{\mathrm{C}}\left(t\right)S_t-f_t,$    is riskless. From \eqref{GrindEQ__13_} and \eqref{GrindEQ__14_}, $\aleph $'s portfolio dynamics in $t\in [0,T]$ is determined by $\ dP^{\left({\mathrm{C}}_{\mathrm{t}}\right)}_t\mathrm{=}{\mathrm{\Delta }}_{\mathrm{C}}\left(t\right)dS_t-df_t=r_tP^{\left(\mathrm{C}\right)}_tdt+dU^{\left(\aleph ,\gamma \right)}_t,\ \ $where
\begin{equation} 
\label{GrindEQ__15_} 
	dU^{\left(\aleph ,\gamma \right)}_t=\frac{{\mu }_t}{2{\mathbb{R}}_{\tau }{\sigma }^2_t}\left(\left({\mu }_t-r_t\right)dt+{\sigma }_tdB\left(t\right)\right)\ ,U^{\left(\aleph ,\gamma \right)}_T=0 
\end{equation} 
is the instantaneous unhedged portfolio part of ${\aleph}$'s portfolio, and thus, $dU^{\left(\aleph ,\gamma \right)}_t$ to $0$ as $t\uparrow T.$ 

\section{Option Pricing with Optimal Quadratic Utility when the Underlying Asset Price follows Continuous Diffusion with Stochastic Volatility}

\noindent Consider now Heston's stochastic volatility model \citep[see][]{Heston:1993} for the underlying asset dynamics,
\begin{equation} \label{GrindEQ__16_} 
	{dS}_t={\mu }_tS_tdt+h\left(v_t\right)S_tdB\left(t\right),\ S_0>0, t\in \left[0,T\right],
\end{equation} 
where 
\begin{enumerate}[label=\alph*)]
	\item the asset volatility driving process $v_t,\ \ t\in \left[0,T\right]\ $ is determined by 
	\begin{equation}
	\label{GrindEQ__17_} 
	dv_t={\alpha }_tv_tdt+{\beta }_tv_tdB^{\left(v\right)}\left(t\right),\ v_0>0,\ t\in \left[0,T\right],\ \  
	\end{equation} 
	\item $B\left(t\right)$ and $B^{\left(v\right)}\left(t\right),t\in \left[0,T\right]$ are correlated Brownian motions, $dB\left(t\right)dB^{\left(v\right)}\left(t\right)=\rho dt,\rho \in \left(-1,1\right),$ generating a stochastic basis $\left(\mathrm{\Omega },\mathbb{F}=\left({\mathcal{F}}_t,\ t\in \left[0,T\right]\right)\mathrm{,}\mathrm{\ }\mathbb{P}\right)\mathrm{;}$
	\item the function $h\left(x\right),x>0,$ is strictly increasing and sufficiently smooth.\footnote{As example for $h\left(x\right),x>0$, one can consider $\ h\left(x\right)=x^a,\ a\in \left(0,1\right)$, and   $h\left(x\right)=lnx$.} 
\end{enumerate}

\noindent      We assume that the market consists of the following assets available for trade:
\begin{enumerate}[label=(\roman*)]
	\item Asset (denoted as $\mathbb{S}$), with price dynamics, $S_t, t\in \left[0,T\right],\ $given by \eqref{GrindEQ__16_};
	\item Asset volatility (denoted as $\mathbb{V}$),\footnote{Examples of market traded $\mathbb{V}$ are: Cboe Equity VIX on Apple (VXAPL), on Amazon (VXAZN), on Goldman Sachs (VXGS), on IBM(VXIBM).} with price dynamics  $v_t,\ \ \ t\in \left[0,T\right],\ $given by \eqref{GrindEQ__17_};
	\item Riskless asset (denoted by $\mathbb{B}$), with price dynamics${\ \beta }_t,\ \ \ t\in \left[0,T\right]$ given by
	\begin{equation} \label{GrindEQ__18_} 
	d{\beta }_t=r_td{\beta }_t,{\beta }_0>0,  t\in \left[0,T\right], 
	\end{equation} 
	where $r_t>0$ is the riskless rate with sup$\mathrm{\{}$ $r_t+\frac{1}{r_t},\ t\in \left[0,T\right]\}<\infty ,$ $\mathbb{P}\mathrm{-}\ a.s.$ 
	\item A ECC ( a derivative, denoted by $\mathbb{C}$) with price process, ${\mathcal{C}}_t=\mathcal{C}\left(S_t,v_t,t\right),\ \ \ t\in \left[0,T\right]$, where the function $\mathcal{C}\left(x,y,t\right),x>0,y>0,\ t\in \left[0,T\right]$  has continues derivatives $\frac{\partial \mathcal{C}\left(x,y,t\right)}{\partial t},\ \frac{{\partial }^2\mathcal{C}\left(x,y,t\right)}{\partial x^2},$ $\frac{{\partial }^2\mathcal{C}\left(x,y,t\right)}{\partial x\partial y},\frac{{\partial }^2\mathcal{C}\left(x,y,t\right)}{\partial y^2}$. 
\end{enumerate}

We assume the model parameters in \eqref{GrindEQ__14_}, \eqref{GrindEQ__15_}, and \eqref{GrindEQ__16_} guarantee the market $(\mathbb{S},\mathbb{V},\mathbb{B})$ is free of any arbitrage and is complete. That is, $(\mathbb{S},\mathbb{V},\mathbb{B})$  (and so does $(\mathbb{S},\mathbb{V},\mathbb{B}\mathrm{,}\mathbb{C}\mathrm{)}\mathrm{)}$   admits unique equivalent martingale measure $\mathbb{Q}$.\footnote{See \citet[Section 6.I, p.118]{Duffie:2001} for sufficient conditions implying the existence and uniqueness of an equivalent measure $\mathbb{Q}$.}

Suppose now the  $\aleph $ is taking a short position in $\mathbb{C}$ having available assets $\mathbb{S},\mathbb{V},\mathbb{B}$ to trade in the self-financing instantaneously riskless portfolio (denoted by ${\mathbb{P}}^{\left(ir\right)}$), with price process $P^{(ir)}_t,\ t\in \left[0,T\right]$   
\begin{align*} 
	P^{(ir)}_t=a^{\left(ir\right)}_tS_t+b^{\left(ir\right)}_tv_t-{\mathcal{C}}_t. 
\end{align*} 
Then, $dP^{\left(ir\right)}_t=a^{\left(ir\right)}_tdS_t+b^{\left(ir\right)}_tdv_t-d{\mathcal{C}}_t=r_tP^{\left(ir\right)}_tdt\ $and therefore the holdings in $\mathbb{S}$ and $\mathbb{V},$  are determined by\footnote{See \citet[Section 5.I]{Duffie:2001}.}
$a^{\left(ir\right)}_t=\frac{\partial \mathcal{C}\left(S_t,v_t,t\right)}{\partial x}$, and  $b^{\left(ir\right)}_t=\frac{\partial\mathcal{C}\left(S_t,v_t,t\right)}{\partial y}.$                    

\noindent The BSM PDE for $\mathcal{C}\left(x,y,t\right),\ x>0,y>0,t\in [0,T)$ is given by

\[\frac{\partial \mathcal{C}\left(x,y,t\right)}{\partial t}+r_t\frac{\partial \mathcal{C}\left(x,y,t\right)}{\partial x}x-r_t\frac{\partial \mathcal{C}\left(x,y,t\right)}{\partial y}y-r_t\mathcal{C}\left(x,y,t\right)+\] 
\[+\frac{1}{2}\frac{{\partial }^2\mathcal{C}\left(x,y,t\right)}{\partial x^2}h{\left(y\right)}^2x^2+\frac{{\partial }^2\mathcal{C}\left(x,y,t\right)}{\partial x\partial y}\rho h\left(y\right){\beta }_txy+\frac{1}{2}\frac{{\partial }^2\mathcal{C}\left(x,y,t\right)}{\partial y^2}{\beta }^2_ty^2=0.\] 

Next, suppose  $\aleph $ would like to determine the optimal holdings in $\mathbb{S}\ $ and $\mathbb{V}$ , denoted respectively as $a^{\left(C_t\right)}_t$ and $\ b^{\left(C_t\right)}_t,\ C_t\in \left(0,1\right)$, which will guarantee that the portfolio value 

\noindent $P^{(C_t)}_t=a^{\left(C_t\right)}_tS_t+b^{\left(C_t\right)}_tv_t-{\mathcal{C}}_t$     maximizes the instantaneous  quadratic utility   ${\mathcal{U}}_t\left(dP^{\left(C_t\right)}_t\right)$ determined by  ${\mathcal{U}}_t\left(dP^{\left(C_t\right)}_t\right)dt=\left(1-C_t\right){\mathbb{E}}_t(dP^{\left(C_t\right)}_t)-C_tvar_t({dP}^{\left(C_t\right)}_t)$. This leads to 
\[a^{\left(C_t\right)}_t=\frac{\partial \mathcal{C}\left(S_t,v_t,t\right)}{\partial x}+\left(\frac{1}{1-{\rho }^2}\right)\frac{{\mu }_t}{2{\mathbb{R}}_t{h{\left(v_t\right)}^2S}_t}-\left(\frac{\rho }{1-{\rho }^2}\right)\frac{{\alpha }_t}{2{\mathbb{R}}_t{\beta }_th\left(v_t\right)S_t},\ \] 
and
\[b^{\left(C_t\right)}_t=\frac{\partial \mathcal{C}\left(S_t,v_t,t\right)}{\partial y}+\left(\frac{1}{1-{\rho }^2}\right)\frac{{\alpha }_t}{2\ {\mathbb{R}}_t{\beta }^2_tv_t}-\left(\frac{\rho }{1-{\rho }^2}\right)\frac{{\mu }_t}{2\ {\mathbb{R}}_th\left(v_t\right){\beta }_tv_t},\] 
with ${\mathbb{R}}_t=\frac{{\mathrm{C}}_{\mathrm{t}}}{\mathrm{1-}{\mathrm{C}}_{\mathrm{t}}}$.

\noindent The value of $\aleph $`s portfolio becomes
\begin{equation} \label{GrindEQ__19_} 
	P^{(C)}_t=a^{\left(C\right)}_tS_t+b^{\left(C\right)}_tv_t-{\mathcal{C}}_t=a^{\left(ir\right)}_tS_t+b^{\left(ir\right)}_tv_t-\left({\mathcal{C}}_t-{\mathcal{C}}^{\left(Risk\ Prem\right)}_t\right), 
\end{equation} 
where $a^{\left(ir\right)}_t$ and $b^{\left(ir\right)}_t$, given by \eqref{GrindEQ__18_}, are the delta-positions in the riskless portfolio, and the risk premium ${\mathcal{C}}^{\left(Risk\ Prem\right)}_t$ is given by
\begin{equation}\label{GrindEQ__20_} 
{\mathcal{C}}^{\left(Risk\ Prem\right)}_t\coloneqq \frac{1}{2{\mathbb{R}}_t\left(1-{\rho }^2\right)}\left(\frac{{\mu }_t}{h{\left(v_t\right)}^2}+\frac{{\alpha }_t}{{\beta }^{2\ }_t}-\rho \frac{{\mu }_t+{\alpha }_t}{h\left(v_t\right){\beta }_t}\right). 
\end{equation}

As in Section4, $\aleph \ $chooses  ${\mathbb{R}}_t\in (0,\infty )$ with ${{\mathrm{lim}}_{t\uparrow T} {\mathbb{R}}_t\ }=\infty $, so that the risk premium ${\mathcal{C}}^{\left(Risk\ Prem\right)}_t$vanishes as $t$ approaches the maturity time $T$, guaranteeing that $\aleph $`s portfolio

\noindent $P^{(C_t)}_t=a^{\left(C_t\right)}_tS_t+b^{\left(C_t\right)}_tv_t-{\mathcal{C}}_t\ $ is riskless as $t\uparrow T.$

\section{Option Pricing with Optimal Quadratic Utility when the Underlying Asset Price follows Continuous Diffusion with Stochastic Volatility and Volatility-of-Volatility}

\noindent A recent extension of Heston's stochastic volatility model is the stochastic volatility and volatility-of-volatility (vol-of-vol) model.\footnote{See \cite{Drimus:2011}, \cite{Gao:2017}, \cite{Huang:2018}, \cite{Branger:2018}, and \cite{Sueppel:2018}.} The Chicago Board Options Exchange introduced the CBOE VIX of VIX Index (CBOE VVIX) as a volatility-of-volatility measure representing the expected volatility of the 30-day forward price of the CBOE Volatility Index (the VIX).\footnote{\ See\ the\ CBOE\ White\ Paper\ ``Double\ the\ Fun\ with\ CBOE's\ VVIX''\ available\ at\ $  $http://www.cboe.com/products/vix-index-volatility/volatility-on-asset-indexes/the-cboe-vvix-index/vvix-whitepaper.\ } The price dynamics under the vol-of-vol model is given by
\begin{equation} \label{GrindEQ__21_} 
	{dS}_t={\mu }_tS_tdt+h\left(v_t\right)S_tdB\left(t\right),\ S_0>0,\ t\in \left[0,T\right],
\end{equation} 
where

\begin{enumerate}[label=\alph*)]
	\item the asset volatility driving process $v_t, \ t\in \left[0,T\right]\ $ is determined by 
	\begin{equation} \label{GrindEQ__22_} 
	dv_t={\alpha }_tv_tdt+g(w_t)v_tdB^{\left(v\right)}\left(t\right),\ v_0>0,\ t\in \left[0,T\right],
	\end{equation} 
	\item  the asset vol-of-vol driving process $w_t, \ t\in \left[0,T\right]\ $ is determined by 
	\begin{equation} \label{GrindEQ__23_} 
	dw_t={\gamma }_tw_tdt+{\delta }_tw_tdB^{\left(w\right)}\left(t\right),\ v_0>0,\ t\in \left[0,T\right],  
	\end{equation} 
	\item $B\left(t\right)$, $B^{\left(v\right)}\left(t\right)$, and $B^{\left(w\right)}\left(t\right),$ $t\in \left[0,T\right]$, are correlated Brownian motions, with
	
	\noindent $dB\left(t\right)dB^{\left(v\right)}\left(t\right)={\rho }^{\left(v\right)}dt$, $dB\left(t\right)dB^{\left(w\right)}\left(t\right)={\rho }^{\left(w\right)}dt$, and $dB^{\left(v\right)}\left(t\right)dB^{\left(w\right)}\left(t\right)={\rho }^{\left(v,w\right)}dt$
	
	\noindent for  $\left({\rho }^{\left(v\right)},{\rho }^{\left(w\right)},{\rho }^{\left(v,w\right)}\right)\ \in {\left(-1,1\right)}^3$. The Brownian motions, $B\left(t\right),B^{\left(v\right)}\left(t\right)$ and $B^{\left(w\right)}\left(t\right),t\in \left[0,T\right]$ generate a stochastic basis $\left(\mathrm{\Omega },\mathbb{F}=\left({\mathcal{F}}_t,\ t\in \left[0,T\right]\right)\mathrm{,}\mathrm{\ }\mathbb{P}\right).$ 
	
	\item the functions $h\left(x\right)$ and $g\left(x\right)$, $x>0,$ are strictly increasing and sufficiently smooth.\footnote{As an example for $h\left(x\right),g\left(x\right),\ x>0$,  one can consider $h\left(x\right)=x^a,\ g\left(x\right)=x^b,\ (a,b)\in {\left(0,1\right)}^2$, and $h\left(x\right)=g(x)=lnx$.} 
\end{enumerate}

\noindent      We assume that the market consists of the following assets available for trade:

\begin{enumerate}[label=(\roman*)]
\item 	Asset (denoted as $\mathbb{S}$), with price dynamics, $S_t,\ \ \ t\in \left[0,T\right],\ $given by \eqref{GrindEQ__21_};\footnote{In our vol-of-vol model we choose the SPDR S\&P 500 ETF (SPY) as an example for the market traded security $(\mathbb{S})$. See  \url{https://www.morningstar.com/etfs/arcx/spy/quote.html}. }
\item  Asset Volatility (denoted as $\mathbb{V}$)\footnote{We choose the CBOE Equity VIX (the volatility index for SPY), as example of market traded volatility  $(\mathbb{V})$. See   \url{http://www.cboe.com/vix}. }, with price dynamics  $v_t,\ t\in \left[0,T\right]$, given by \eqref{GrindEQ__22_};
\item Asset Vol-of-Vol (denoted as $\mathbb{W}$)\footnote{We choose as an example of market traded vol-of-vol $(\mathbb{W})$, the CBOE VIX volatility index (VVIX). See \url{https://finance.yahoo.com/quote/\%5Evvix?ltr=1}.}, with price dynamics  $w_t, \ t\in \left[0,T\right]$, given by \eqref{GrindEQ__23_};
\item Riskless asset (denoted by $\mathbb{B}$), with price dynamics$\ {\beta }_t, t\in \left[0,T\right]$ given by \eqref{GrindEQ__16_}.
\item A ECC (a derivative, denoted by $\mathbb{C}$) with price process, ${\mathcal{C}}_t=\mathcal{C}\left(S_t,v_t,w_t,\ t\right),\ \ \ t\ \ \left[0,T\right]$, where the function $\mathcal{C}\left(x,y,z,t\right),x>0,y>0,z>0,\ t\in \left[0,T\right]$  has continues $\frac{\partial \mathcal{C}\left(x,y,t\right)}{\partial t},\ $and continuous derivatives of 3${}^{rd}$ order with respect to $,x>0,y>0,z>0.$
\end{enumerate}

We assume the model parameters in \eqref{GrindEQ__21_}, \eqref{GrindEQ__22_}, \eqref{GrindEQ__23_}, and \eqref{GrindEQ__16_} guarantee the market $(\mathbb{S},\mathbb{V},\mathbb{W},\mathbb{B})$ is without free of any arbitrage opportunities and complete; that is, $(\mathbb{S},\mathbb{V},\mathbb{W},,\mathbb{B})$  (and so does $(\mathbb{S},\mathbb{V},\mathbb{W},,\mathbb{B}\mathrm{,}\mathbb{C}\mathrm{)}$   admits unique equivalent martingale measure $\mathbb{Q}.$

Suppose now $\aleph $ is taking a short position in $\mathbb{C}$ having available assets $\mathbb{S},\mathbb{V},\mathbb{W},\mathbb{B}$ to trade in the self-financing instantaneously riskless portfolio (denoted by ${\mathbb{P}}^{\left(ir\right)}$), with price process $P^{(ir)}_t,\ t\in \left[0,T\right]$   
\begin{equation} \label{GrindEQ__24_} 
	P^{(ir)}_t=a^{\left(ir\right)}_tS_t+b^{\left(ir\right)}_tv_t+c^{\left(ir\right)}_tw_t-{\mathcal{C}}_t 
\end{equation} 
implying,  $dP^{\left(ir\right)}_t=a^{\left(ir\right)}_tdS_t+b^{\left(ir\right)}_tdv_t+c^{\left(ir\right)}_tdw_t-d{\mathcal{C}}_t=r_tP^{\left(ir\right)}_tdt\ $and thus the holdings in $\mathbb{S}$, $\mathbb{V},$  and $\mathbb{W}$ , are determined by 
\begin{equation}\label{GrindEQ__25_}
a^{\left(ir\right)}_t=\frac{\partial \mathcal{C}\left(S_t,v_t,w_t,t\right)}{\partial x},   b^{\left(ir\right)}_t=\frac{\partial \mathcal{C}\left(S_t,v_t,w_t,t\right)}{\partial y}, \, \text{and}\,\,        c^{\left(ir\right)}_t=\frac{\partial \mathcal{C}\left(S_t,v_t,w_t,t\right)}{\partial z}.  
\end{equation}

The BSM PDE for  $\mathcal{C}\left(x,y,z,t\right),x>0,y>0,z>0,\ t\in \left[0,T\right),$ is given by\footnote{See \citet[Section 5.I]{Duffie:2001}.} 
\begin{equation}
\label{GrindEQ__26_}
\begin{array}{ccc}
\frac{\partial \mathcal{C}\left(x,y,z,\ t\right)}{\partial t}+r_t\frac{\partial \mathcal{C}\left(x,y,z,\ t\right)}{\partial x}x+r_t\frac{\partial \mathcal{C}\left(x,y,z,\ t\right)}{\partial y}y+r_t\frac{\partial \mathcal{C}\left(x,y,z,\ t\right)}{\partial z}z-r_t\mathcal{C}\left(x,y,z,\ t\right)\\
+\frac{1}{2}\frac{{\partial }^2\mathcal{C}\left(x,y,z,\ t\right)}{\partial x^2}h{\left(y\right)}^2x^2+\frac{1}{2}\frac{{\partial }^2\mathcal{C}\left(x,y,z,\ t\right)}{\partial y^2}g{\left(z\right)}^2y^2
+\frac{1}{2}\frac{{\partial }^2\mathcal{C}\left(x,y,z,\ t\right)}{\partial z^2}{\delta }^2_tz^2\\
+{\rho }^{\left(v\right)}\frac{{\partial }^2\mathcal{C}\left(x,y,z,\ t\right)}{\partial x\partial y}h\left(y\right)g\left(z\right)xy
+{\rho }^{\left(w\right)}\frac{{\partial }^2\mathcal{C}\left(x,y,z,\ t\right)}{\partial x\partial z}h\left(y\right){\delta }_txz+{\rho }^{\left(v,w\right)}\frac{{\partial }^2\mathcal{C}\left(x,y,z,\ t\right)}{\partial y\partial z}g\left(z\right){\delta }_tyz=0.

\end{array}
\end{equation}
with boundary $\mathcal{C}\left(x,y,z,T\right)=G\left(x,y,z\right),$ where $G\left(S_T,v_T,w_T\right)$ is the terminal payoff of derivative $\mathbb{C}$. The Feynman-Kac probabilistic solution of \eqref{GrindEQ__26_} is given in Appendix E in \cite{Duffie:2001}.

\noindent Next, suppose  $\aleph $ would like to determine the optimal holdings in $\mathbb{S}\ $,$\mathbb{V}$, and $\mathbb{W}$ denoted respectively as $a^{\left(C_t\right)}_t$, $\ b^{\left(C_t\right)}_t,$ and $\ c^{\left(C_t\right)}_t$ $,\ C_t\in \left(0,1\right),\ t\left[0,T\right]$, which will guarantee that the portfolio value  $P^{\left(C_t\right)}_t=a^{\left(C_t\right)}_tS_t+b^{\left(C_t\right)}_tv_t+c^{\left(C_t\right)}_tw_t-{\mathcal{C}}_t$     maximizes the instantaneous quadratic utility     ${\mathcal{U}}_t\left(dP^{\left(C_t\right)}_t\right)$ determined by ${\mathcal{U}}_t\left(dP^{\left(C_t\right)}_t\right)dt=\left(1-C_t\right){\mathbb{E}}_t(dP^{\left(C_t\right)}_t)-C_tvar_t({dP}^{\left(C_t\right)}_t)$. This leads to
$a^{\left(C_t\right)}_t=a^{\left(ir\right)}_t+\frac{1}{2{\mathbb{R}}_th\left(v_t\right)S_t}\frac{D_a}{D}$,  $b^{\left(C\right)}_t=b^{\left(ir\right)}_t+\frac{1}{2{\mathbb{R}}_tg\left(w_t\right)v_t}\frac{D_b}{D}$, and $c^{\left(C\right)}_t=c^{\left(ir\right)}_t+\frac{1}{2{\mathbb{R}}_t{\delta }_tw_t}\frac{D_c}{D},$
 where $D$, $D_a$, $D_b$, and $D_c$ are the following $3\times 3$-determinants
\[D=\left| \begin{array}{ccc}
1 & {\rho }^{\left(v\right)} & {\rho }^{\left(w\right)} \\ 
{\rho }^{\left(v\right)} & 1 & {\rho }^{\left(v,w\right)} \\ 
{\rho }^{\left(w\right)} & {\rho }^{\left(v,w\right)} & 1 \end{array}
\right|,D_a=\left| \begin{array}{ccc}
\frac{{\mu }_t}{h\left(v_t\right)} & {\rho }^{\left(v\right)} & {\rho }^{\left(w\right)} \\ 
\frac{{\alpha }_t}{g\left(w_t\right)} & 1 & {\rho }^{\left(v,w\right)} \\ 
\frac{{\gamma }_t}{{\delta }_t} & {\rho }^{\left(v,w\right)} & 1 \end{array}
\right|,\] 
\[D_b=\left| \begin{array}{ccc}
1 & \frac{{\mu }_t}{h\left(v_t\right)} & {\rho }^{\left(w\right)} \\ 
{\rho }^{\left(v\right)} & \frac{{\alpha }_t}{g\left(w_t\right)} & {\rho }^{\left(v,w\right)} \\ 
{\rho }^{\left(w\right)} & \frac{{\gamma }_t}{{\delta }_t} & 1 \end{array}
\right|,D_c=\left| \begin{array}{ccc}
1 & {\rho }^{\left(v\right)} & \frac{\left(1-C_t\right){\mu }_t}{h\left(v_t\right)} \\ 
{\rho }^{\left(v\right)} & 1 & \frac{{\alpha }_t}{g\left(w_t\right)} \\ 
{\rho }^{\left(w\right)} & {\rho }^{\left(v,w\right)} & \frac{{\gamma }_t}{{\delta }_t} \end{array}
\right|.\] 
$\aleph$ chooses  ${\mathbb{R}}_t\in \left(0,\infty \right)$ with ${{\mathrm{lim}}_{t\uparrow T} {\mathbb{R}}_t\ }=\infty $, so that the delta positions $(a^{\left(C_t\right)}_t,b^{\left(C_t\right)}_t,c^{\left(C_t\right)}_t)$
maximizing the instantaneous utility ${\mathcal{U}}_t\left(dP^{\left(C_t\right)}_t\right)\ $approaches (as $t\uparrow T$)  the risk-neutral delta-positions $(a^{\left(ir\right)}_t,b^{\left(ir\right)}_t,c^{\left(ir\right)}_t)$ $T$. Thus, $\aleph $`s portfolio $P^{(C_t)}_t=a^{\left(C_t\right)}_tS_t+b^{\left(C_t\right)}_tv_t+c^{\left(C_t\right)}_tw_t-{\mathcal{C}}_t\ $ is asymptotically riskless as $t\uparrow T.$

\section{Option Pricing with Optimal Quadratic Utility when the Underlying Asset Price follows Jump-Diffusion Process}

\noindent Consider now \cite{Merton:1976} \footnote{See also \cite{Runggaldier:2003}, and \cite{Rachev:2017}.} jump diffusion model with three assets ${(\mathbb{S}}^{\left(1\right)},{\mathbb{S}}^{\left(2\right)},\ \mathbb{B})\ .$ The riskless asset (denoted by $\mathbb{B}$) has price dynamics$\ {\beta }_t,\ t\in \left[0,T\right]$ given by \eqref{GrindEQ__18_}. The two risky assets ${(\mathbb{S}}^{\left(1\right)},{\mathbb{S}}^{\left(2\right)})\ $have price processes $S^{\left(j\right)}_t\ ,t\in \left[0,T\right],j=1,2,$ with jump-diffusion dynamics 
\begin{equation}
\label{GrindEQ__27_}
\frac{dS^{\left(j\right)}_t}{S^{\left(j\right)}_t}={\mu }^{\left(j\right)}_tdt+{\sigma }^{\left(j\right)}_tdB\left(t\right)+{\gamma }^{\left(j\right)}_tdN\left(t\right),t\in \left[0,T\right],\ {\ \ S}^{\left(j\right)}_0>0,\ \ \ \ j=1,2,
\end{equation}
where ${\mu }^{\left(j\right)}_t\in R,{\sigma }^{\left(j\right)}_t>0,\ \ {\gamma }^{\left(j\right)}_t\in R $  Because the dynamics of both risky assets ${\mathbb{S}}^{\left(1\right)},{\mathbb{S}}^{\left(2\right)}$ are driven by the same pair of random processes $\left(B\left(t\right),N\left(t\right)\right)$,  $t\in \left[0,T\right]$, one can view ${\mathbb{S}}^{\left(2\right)}$ as an ECC with underlying asset  ${\mathbb{S}}^{\left(1\right)}$. $T_2$, the maturity of ${\mathbb{S}}^{\left(2\right)}$, is assumed to be greater than $T$. The triplet $\left(S^{\left(1\right)}_t,S^{\left(2\right)}_t,{\beta }_t\right),\ \ t\in \left[0,T\right]$ is defined on a stochastic basis $\left(\mathrm{\Omega },\mathcal{F},\mathbb{F}=\left\{{\mathcal{F}}_t,\ \ t\in \left[0,T\right]\right\}\mathrm{,}\mathbb{P}\right)$, representing the natural world. 

The basis $\left(\mathrm{\Omega },\mathcal{F},\mathbb{F}=\left\{{\mathcal{F}}_t,t\in \left[0,T\right]\right\}\mathrm{,}\mathbb{P}\right)$ is generated by the Brownian motion $B\left(t\right),t\in \left[0,T\right]$ and a non-homogeneous Poisson process $N\left(t\right),t\in \left[0,T\right],$ with intensity ${\lambda }_t>0,t\in \left[0,T\right].$ Denote by $M\left(t\right)=N\left(t\right)-{\lambda }_t,t\in \left[0,T\right],$ the martingale corresponding to $N\left(t\right),t\in \left[0,T\right].$ Under the equivalent martingale measure (EMM) $\mathbb{Q}\mathrm{\sim }\mathbb{P}\mathrm{,}$
\begin{equation} \label{GrindEQ__28_} 
\frac{dS^{\left(j\right)}_t}{S^{\left(j\right)}_t}=r_tdt+{\sigma }^{\left(j\right)}_tdB^{\mathbb{Q}}\left(t\right)+{\gamma }^{\left(j\right)}_tdM^{\mathbb{Q}}\left(t\right),t\in \left[0,T\right], 
\end{equation} 
where $B^{\mathbb{Q}}\left(t\right)$ $,t\in \left[0,T\right],\ $is a Brownian motion on $\mathbb{Q}$ and $M^{\mathbb{Q}}\left(t\right),t\in \left[0,T\right],$ is a Poisson martingale on $\mathbb{Q}.$ On $\mathbb{P},\ $the dynamics of  $B^{\mathbb{Q}}\left(t\right)$ and $M^{\mathbb{Q}}\left(t\right),\ t\in \left[0,T\right],$ is determined by the market-price-of-risk densities ${\vartheta }_t$ and ${\lambda }_t\left(1-{\eta }_t\right),\ t\in \left[0,T\right]$:
\begin{equation}\label{GrindEQ__29_} 
B^{\mathbb{Q}}\left(t\right)=B\left(t\right)+\int^t_0{{\vartheta }_tds},\ \text{and} \,M^{\mathbb{Q}}\left(t\right)=M\left(t\right)+\int^t_0{{\lambda }_s\left(1-{\eta }_s\right)ds},\ \ \ t\in \left[0,T\right].
\end{equation}

Suppose $\aleph $ enters a short position in ECC-contract $\mathbb{V}$ with price process $f_t=V\left(S^{\left(1\right)}_t,S^{\left(2\right)}_t,t\right)$, and terminal value $f_T=V\left(S^{\left(1\right)}_T,S^{\left(2\right)}_T,T\right)=G\left(S^{\left(1\right)}_T,S^{\left(2\right)}_T\right)$. The uniqueness of the EMM $\mathbb{Q}$  is a sufficient condition to obtain a unique arbitrage-free price of $\mathbb{V}$ as $f_t={\mathbb{E}}^{\left(\mathbb{Q}\right)}\left(e^{-\int^T_t{r_sds}}G\left(S^{\left(1\right)}_T,S^{\left(2\right)}_T\right)\slash {\mathcal{F}}_t\right)$.\footnote{See \citet[p. 202]{Runggaldier:2003}} 

Now $\aleph $ forms a self-financing portfolio $P^{\left({\mathrm{C}}_{\mathrm{t}}\right)}_t\mathrm{=}{{\mathrm{\Delta }}_{\mathrm{C}}}^{\left(1\right)}_t\left(t\right)S^{\left(1\right)}_t+{{\mathrm{\Delta }}_{\mathrm{C}}}^{\left(2\right)}_t\left(t\right)S^{\left(2\right)}_t-f_t,\ $ where $f_t=V\left(S^{\left(1\right)}_t,S^{\left(2\right)}_t,t\right)\ $is the fair price of the option at time $t\in \left\{0,T\right).$ The function $V\left(x_1,x_2,t\right),\ x_1>0,x_2>0,t\in \left[0,T\right)$ is sufficiently smooth and thus, by the It\^{o} formula for jump-diffusions, $f_t=V\left(S^{\left(1\right)}_t,S^{\left(2\right)}_t,t\right)\ $is again a jump-diffusion process (see \cite{Carr:2007}) with price dynamics given by

\begin{equation}
\label{GrindEQ__30_}
\begin{array}{lll}
df_t&=
\left( \frac{\partial V\left(S^{\left(1\right)}_t,S^{\left(2\right)}_t,t\right)}{\partial t}+{\mu }^{\left(1\right)}_t\frac{\partial V\left(S^{\left(1\right)}_t,S^{\left(2\right)}_t,t\right)}{\partial x_1}S^{\left(1\right)}\left(t\right)+{\mu }^{\left(2\right)}_t\frac{\partial V\left(S^{\left(1\right)}_t,S^{\left(2\right)}_t,t\right)}{\partial x_2}S^{\left(2\right)}\left(t\right)\right) dt\\
&+\left( \frac{1}{2}\frac{\partial V^2\left(S^{\left(1\right)}_t,S^{\left(2\right)}_t,t\right)}{\partial x^2_1}{\left(S^{\left(1\right)}\left(t\right)\right)}^2{\left({\sigma }^{\left(1\right)}_t\right)}^2+\frac{1}{2}\frac{\partial V^2\left(S^{\left(1\right)}_t,S^{\left(2\right)}_t,t\right)}{\partial x^2_2}{\left(S^{\left(2\right)}\left(t\right)\right)}^2{\left({\sigma }^{\left(2\right)}_t\right)}^2\right) dt\\
&+\left( \frac{{\partial }^2V\left(S^{\left(1\right)}_t,S^{\left(2\right)}_t,t\right)}{\partial x_1\partial x_2}+S^{\left(1\right)}\left(t\right)S^{\left(2\right)}\left(t\right){\sigma }^{\left(2\right)}_t{\sigma }^{\left(2\right)}_t\right) dt
\\
&+\left(  {\sigma }^{\left(1\right)}_t\frac{\partial V\left(S^{\left(1\right)}_t,S^{\left(2\right)}_t,t\right)}{\partial x_1}S^{\left(1\right)}\left(t\right)+{\sigma }^{\left(2\right)}_t\frac{\partial V\left(S^{\left(1\right)}_t,S^{\left(2\right)}_t,t\right)}{\partial x_2}S^{\left(2\right)}\left(t\right)\right)   dB\left(t\right)\\
&+\left( V\left(S^{\left(1\right)}_t+{\gamma }^{\left(1\right)}_t,S^{\left(2\right)}_t+{\gamma }^{\left(2\right)}_t,t\right)-V\left(S^{\left(1\right)}_t,S^{\left(2\right)}_t,t\right)\right) dN\left(t\right).
\end{array}
\end{equation}

Suppose first that  $\aleph $ would like to find the delta-positions $\left({{\mathrm{\Delta }}_{\mathrm{C}}}^{\left(rn,1\right)}_t\left(t\right),{{\mathrm{\Delta }}_{\mathrm{C}}}^{\left(rn,2\right)}_t\left(t\right)\right)\ $that will guarantee the perfect hedge of the short position in the derivatives; that is,  $\aleph $'s  portfolio $P^{\left(rn\right)}_t\mathrm{=}{{\mathrm{\Delta }}_{\mathrm{C}}}^{\left(rn,1\right)}_t\left(t\right)S^{\left(1\right)}_t+{{\mathrm{\Delta }}_{\mathrm{C}}}^{\left(2\right)}_t\left(t\right)S^{\left(rn1\right)}_t-f_t,\ t\in \left[0,T\right]$ is instantaneously riskless,

\begin{equation}
\label{GrindEQ__31_}
dP^{\left(rn\right)}_t={{\mathrm{\Delta }}_{\mathrm{C}}}^{\left(rn,1\right)}_t\left(t\right)dS^{\left(1\right)}_t+{{\mathrm{\Delta }}_{\mathrm{C}}}^{\left(rn,2\right)}_t\left(t\right)dS^{\left(2\right)}_t-df_t=r_tP^{\left({rn,C}_t\right)}_tdt.
\end{equation}

Then \eqref{GrindEQ__30_} and \eqref{GrindEQ__31_} imply that risk-neutral delta positions ${{\mathrm{\Delta }}_{\mathrm{C}}}^{\left(rn,1\right)}_t\left(t\right)$ and ${{\mathrm{\Delta }}_{\mathrm{C}}}^{\left(rn,2\right)}_t\left(t\right)$ are determined by 
\[{{\mathrm{\Delta }}_C}^{(rn,1)}_t\left(t\right)=\frac{\left( \begin{array}{c}
	{\sigma }^{\left(1\right)}_t\frac{\partial V\left(S^{\left(1\right)}_t,S^{\left(2\right)}_t,t\right)}{\partial x_1}S^{\left(1\right)}\left(t\right){\gamma }^{\left(2\right)}_t+{\sigma }^{\left(2\right)}_t\frac{\partial V\left(S^{\left(1\right)}_t,S^{\left(2\right)}_t,t\right)}{\partial x_2}S^{\left(2\right)}\left(t\right){\gamma }^{\left(2\right)}_t\\ 
	-V\left(S^{\left(1\right)}_t+{\gamma }^{\left(1\right)}_t,S^{\left(2\right)}_t+{\gamma }^{\left(2\right)}_t,t\right){\sigma }^{\left(2\right)}_t+V\left(S^{\left(1\right)}_t,S^{\left(2\right)}_t,t\right){\sigma }^{\left(2\right)}_t \end{array}
	\right)}{S^{\left(1\right)}_t\left({\sigma }^{\left(1\right)}_t{\gamma }^{\left(2\right)}_t-{\gamma }^{\left(1\right)}_t{\sigma }^{\left(2\right)}_t\right)},\] 
and
\[{{\mathrm{\Delta }}_C}^{(rn,2)}_t\left(t\right)=\frac{\left( \begin{array}{c}
	V\left(S^{\left(1\right)}_t+{\gamma }^{\left(1\right)}_t,S^{\left(2\right)}_t+{\gamma }^{\left(2\right)}_t,t\right){\sigma }^{\left(1\right)}_t-V\left(S^{\left(1\right)}_t,S^{\left(2\right)}_t,t\right){\sigma }^{\left(1\right)}_t \\ 
	-{\sigma }^{\left(1\right)}_t\frac{\partial V\left(S^{\left(1\right)}_t,S^{\left(2\right)}_t,t\right)}{\partial x_1}S^{\left(1\right)}\left(t\right){\gamma }^{\left(1\right)}_t-{\sigma }^{\left(2\right)}_t\frac{\partial V\left(S^{\left(1\right)}_t,S^{\left(2\right)}_t,t\right)}{\partial x_2}S^{\left(2\right)}\left(t\right){\gamma }^{\left(1\right)}_t \end{array}
	\right)}{S^{\left(2\right)}_t\left({\sigma }^{\left(1\right)}_t{\gamma }^{\left(2\right)}_t-{\sigma }^{\left(2\right)}_t{\gamma }^{\left(1\right)}_t\right)}.\] 

Next, suppose$\ \aleph \ $choses a dynamic relative risk averse parameter $C_t\in (0,1]$, and would like to maximize the conditional instantaneous expected utility function ${\mathcal{U}}_t\left(dP^{\left(C_t\right)}_t\right),$ determined by ${\mathcal{U}}_t\left(dP^{\left(C_t\right)}_t\right)dt=\left(1-C_t\right){\mathbb{E}}_t(dP^{\left(C_t\right)}_t)-C_tvar_t({dP}^{\left(C_t\right)}_t)$, where $dP^{\left(C_t\right)}_t={{\mathrm{\Delta }}_{\mathrm{C}}}^{\left(1\right)}_t\left(t\right)dS^{\left(1\right)}_t+{{\mathrm{\Delta }}_{\mathrm{C}}}^{\left(2\right)}_t\left(t\right)dS^{\left(1\right)}_t-df_t$ $.$ From \eqref{GrindEQ__27_} and \eqref{GrindEQ__30_}, it follows that

\begin{equation*}{}
\begin{array}{lll}
dP^{\left(C_t\right)}_t
={{\mathrm{\Delta }}_{\mathrm{C}}}^{\left(1\right)}_t\left(t\right)dS^{\left(1\right)}_t+{{\mathrm{\Delta }}_{\mathrm{C}}}^{\left(2\right)}_t\left(t\right)dS^{\left(2\right)}_t-df_t\\=
\left( {{\mathrm{\Delta }}_{\mathrm{C}}}^{\left(1\right)}_t\left(t\right)S^{\left(1\right)}_t{\mu }^{\left(1\right)}_t+{{\mathrm{\Delta }}_{\mathrm{C}}}^{\left(2\right)}_t\left(t\right)S^{\left(2\right)}_{t-}{\mu }^{\left(2\right)}_t \right) dt
\\-\left(\frac{\partial V\left(S^{\left(1\right)}_t,S^{\left(2\right)}_t,t\right)}{\partial t}+{\mu }^{\left(1\right)}_t\frac{\partial V\left(S^{\left(1\right)}_t,S^{\left(2\right)}_t,t\right)}{\partial x_1}S^{\left(1\right)}\left(t\right)+{\mu }^{\left(2\right)}_t\frac{\partial V\left(S^{\left(1\right)}_t,S^{\left(2\right)}_t,t\right)}{\partial x_2}S^{\left(2\right)}\left(t\right) \right) dt
\\
-\left( \frac{1}{2}\frac{\partial V^2\left(S^{\left(1\right)}_t,S^{\left(2\right)}_t,t\right)}{\partial x^2_1}{\left(S^{\left(1\right)}\left(t\right)\right)}^2{\left({\sigma }^{\left(1\right)}_t\right)}^2+\frac{1}{2}\frac{\partial V^2\left(S^{\left(1\right)}_t,S^{\left(2\right)}_t,t\right)}{\partial x^2_2}{\left(S^{\left(2\right)}\left(t\right)\right)}^2{\left({\sigma }^{\left(2\right)}_t\right)}^2\right) dt 
\\-\left(\frac{{\partial }^2V\left(S^{\left(1\right)}_t,S^{\left(2\right)}_t,t\right)}{\partial x_1\partial x_2}-S^{\left(1\right)}\left(t\right)S^{\left(2\right)}\left(t\right){\sigma }^{\left(2\right)}_t{\sigma }^{\left(2\right)}_t\right) dt\\
+\left( {{\mathrm{\Delta }}_{\mathrm{C}}}^{\left(1\right)}_t\left(t\right)S^{\left(1\right)}_{t-}{\sigma }^{\left(1\right)}_t+{{\mathrm{\Delta }}_{\mathrm{C}}}^{\left(2\right)}_t\left(t\right)S^{\left(2\right)}_{t-}{\sigma }^{\left(2\right)}_t\right) dB\left(t\right) \\ 
-\left( {\sigma }^{\left(1\right)}_t\frac{\partial V\left(S^{\left(1\right)}_t,S^{\left(2\right)}_t,t\right)}{\partial x_1}S^{\left(1\right)}\left(t\right)+{\sigma }^{\left(2\right)}_t\frac{\partial V\left(S^{\left(1\right)}_t,S^{\left(2\right)}_t,t\right)}{\partial x_2}S^{\left(2\right)}\left(t\right)
\right) dB\left(t\right)
\\
+\left( {{\mathrm{\Delta }}_{\mathrm{C}}}^{\left(1\right)}_t\left(t\right)S^{\left(1\right)}_t{\gamma }^{\left(1\right)}_t+{{\mathrm{\Delta }}_{\mathrm{C}}}^{\left(2\right)}_t\left(t\right)S^{\left(2\right)}_{t-}{\gamma }^{\left(2\right)}_t\right)dN\left(t\right) - \\ 
- \left( V\left(S^{\left(1\right)}_t+{\gamma }^{\left(1\right)}_t,S^{\left(2\right)}_t+{\gamma }^{\left(2\right)}_t,t\right)-V\left(S^{\left(1\right)}_t,S^{\left(2\right)}_t,t\right)  \right)  dN\left(t\right).
\end{array}
\end{equation*}

Thus, from ${\mathcal{U}}_t\left(dP^{\left(C_t\right)}_t\right)dt=\left(1-C_t\right){\mathbb{E}}_t(dP^{\left(C_t\right)}_t)-C_tvar_t({dP}^{\left(C_t\right)}_t)$, we have that the optimal delta positions $\left({{\mathrm{\Delta }}_{\mathrm{C}}}^{\left(1\right)}_t\left(t\right),{{\mathrm{\Delta }}_{\mathrm{C}}}^{\left(2\right)}_t\left(t\right)\right)$ maximizing ${\mathcal{U}}_t\left(dP^{\left(C_t\right)}_t\right)$ are given by 
\[{{\mathrm{\Delta }}_{\mathrm{C}}}^{\left(1\right)}_t\left(t\right)={{\mathrm{\Delta }}_{\mathrm{C}}}^{\left(rn,1\right)}_t\left(t\right)+{\mathcal{E}}^{\left(1\right)}_{C_t}\left(t\right){,\ \ {\mathrm{\Delta }}_{\mathrm{C}}}^{\left(2\right)}_t\left(t\right)={{\mathrm{\Delta }}_{\mathrm{C}}}^{\left(rn,2\right)}_t\left(t\right)+{\mathcal{E}}^{\left(2\right)}_{C_t}\left(t\right),\] 
where 
\[{\mathcal{E}}^{\left(1\right)}_{C_t}\left(t\right)=\frac{1}{{\mathbb{R}}_tS^{\left(1\right)}_t}\left(\frac{{\sigma }^{\left(2\right)}_t\left({\mu }^{\left(1\right)}_t{\sigma }^{\left(2\right)}_t-{\mu }^{\left(2\right)}_t{\sigma }^{\left(1\right)}_t\right)}{{\lambda }_t{\left({\sigma }^{\left(1\right)}_t{\gamma }^{\left(2\right)}_t-{\gamma }^{\left(1\right)}_t{\sigma }^{\left(2\right)}_t\right)}^2}-\frac{{\sigma }^{\left(2\right)}_t}{{\sigma }^{\left(1\right)}_t{\gamma }^{\left(2\right)}_t-{\gamma }^{\left(1\right)}_t{\sigma }^{\left(2\right)}_t}+\frac{{\gamma }^{\left(2\right)}_t\left({\mu }^{\left(1\right)}_t{\gamma }^{\left(2\right)}_t-{\mu }^{\left(2\right)}_t{\gamma }^{\left(1\right)}_t\right)}{{\left({\sigma }^{\left(1\right)}_t{\gamma }^{\left(2\right)}_t-{\gamma }^{\left(1\right)}_t{\sigma }^{\left(2\right)}_t\right)}^2}\right),\ \] 
and
\[{\mathcal{E}}^{\left(2\right)}_{C_t}\left(t\right)=\frac{1}{{\mathbb{R}}_tS^{\left(2\right)}_{t-}}\left( \begin{array}{ccc}
\frac{{\sigma }^{\left(1\right)}_t\left({\mu }^{\left(2\right)}_t{\sigma }^{\left(1\right)}_t-{\mu }^{\left(1\right)}_t{\sigma }^{\left(2\right)}_t\right)}{{\lambda }_t{\left({\gamma }^{\left(1\right)}_t{\sigma }^{\left(2\right)}_t-{\gamma }^{\left(2\right)}_t{\sigma }^{\left(1\right)}_t\right)}^2} \\ 
+\frac{{\lambda }_t{\sigma }^{\left(1\right)}_t\left({\sigma }^{\left(1\right)}_t{\gamma }^{\left(2\right)}_t-{\sigma }^{\left(2\right)}_t{\gamma }^{\left(1\right)}_t\right)}{{\lambda }_t{\left({\gamma }^{\left(1\right)}_t{\sigma }^{\left(2\right)}_t-{\gamma }^{\left(2\right)}_t{\sigma }^{\left(1\right)}_t\right)}^2} \\ 
+\frac{{\lambda }_t{\gamma }^{\left(1\right)}_t\left({\mu }^{\left(2\right)}_t{\gamma }^{\left(1\right)}_t-{\mu }^{\left(1\right)}_t{\gamma }^{\left(2\right)}_t\right)}{{\lambda }_t{\left({\gamma }^{\left(1\right)}_t{\sigma }^{\left(2\right)}_t-{\gamma }^{\left(2\right)}_t{\sigma }^{\left(1\right)}_t\right)}^2} \end{array}
\right),\] 
with ${\mathbb{R}}_t=\frac{{\mathrm{C}}_{\mathrm{t}}}{\mathrm{1-}{\mathrm{C}}_{\mathrm{t}}}\in \left[0,\infty \right).$

The terms ${\mathcal{E}}^{\left(1\right)}_{C_t}\left(t\right)$ and ${\mathcal{E}}^{\left(2\right)}_{C_t}\left(t\right)\ $are expressing the intensity level of unhedged risk at time $t\in \left[0,T\right),$  when $\aleph $ uses as delta positions $\left({{\mathrm{\Delta }}_{\mathrm{C}}}^{\left(1\right)}_t\left(t\right),{{\mathrm{\Delta }}_{\mathrm{C}}}^{\left(2\right)}_t\left(t\right)\right)$ instead of the risk-neutral  deltas $\left({{\mathrm{\Delta }}_{\mathrm{C}}}^{\left(rn,1\right)}_t\left(t\right),{{\mathrm{\Delta }}_{\mathrm{C}}}^{\left(rn,2\right)}_t\left(t\right)\right).\ $As $\aleph $ becomes more and more risk-averse over the period of time $t\in [0,T)$, and ${\mathrm{C}}_{\mathrm{t}}\downarrow 0,{\mathbb{R}}_t\uparrow \infty ,$ as $t\uparrow T,$ the terms ${\mathcal{E}}^{\left(1\right)}_{C_t}\left(t\right)$ and ${\mathcal{E}}^{\left(2\right)}_{C_t}\left(t\right)$ vanish, and $\left({{\mathrm{\Delta }}_{\mathrm{C}}}^{\left(1\right)}_t\left(t\right),{{\mathrm{\Delta }}_{\mathrm{C}}}^{\left(2\right)}_t\left(t\right)\right)\to \left({{\mathrm{\Delta }}_{\mathrm{C}}}^{\left(1\right)}_t\left(t\right),{{\mathrm{\Delta }}_{\mathrm{C}}}^{\left(2\right)}_t\left(t\right)\right),$ $t\uparrow T$ . This guarantees that at the terminal time $T$, $\aleph $ hedges the short position entirely.

\section{Conclusions}

\noindent In this paper, we consider a general model of a trader taking positions in a European contingent claim (ECC) contract and the underlying assets. Between the initiation of the contact and its maturity, the trader forms a portfolio of the underlying assets and a short position in the ECC contract, balancing the desire to optimize the portfolio using the mean-variance framework and hedging the short position in the ECC. When rebalancing the portfolio, the trader continuously updates the risk aversion parameter so that the trader profits from the optimal mean-variance portfolio before the option matures. The mean-variance profit gradually declines and vanishes at the ECC's maturity date on which the trader's portfolio provides a perfect hedge for the short position in the ECC. We examine the trader's optimal allocations in the underlying assets and the amount of unhedged risk in the short ECC-position in five cases: \eqref{GrindEQ__1_} the binomial pricing model, \eqref{GrindEQ__2_} the continuous diffusion pricing model, \eqref{GrindEQ__3_} the stochastic volatility pricing model, \eqref{GrindEQ__4_} the volatility-of-volatility model, and \eqref{GrindEQ__5_} the Merton jump-diffusion model. Our approach can be viewed as a combination of mean-variance analysis and option pricing theory. Our empirical results for the binomial pricing model case indicate that for in-the-money call-option traders with short option positions use the investment opportunity extensively, as suggested in this paper. While for the out-the-money call-option, traders disregard the investment opportunity that the investment strategy offers.

\normalem

\end{spacing}
\end{document}